\documentclass{aa}
\usepackage{graphicx}
\usepackage{times}
\begin{document}

 \title{A three-dimensional model for the radio emission of magnetic chemically 
peculiar stars}
\author{     C. Trigilio\inst{1}
        \and P. Leto\inst{1}	
        \and G. Umana\inst{1}
	\and F. Leone\inst{2}
	\and C.S. Buemi\inst{1}
	}
\institute{ 
		Istituto di Radioastronomia del C.N.R., P.O. Box 141,
		I-96017 Noto (SR), Italy
\and
		INAF - Osservatorio Astrofisico di Catania, Citt\`a Universitaria,
		I-95125 Catania, Italy
	      }

  \offprints{C. Trigilio}
  \mail{c.trigilio@ira.cnr.it}

   \date{Received ; accepted }
   \abstract{
In this paper we present a three-dimensional numerical model for the radio 
emission of Magnetic Chemically Peculiar stars, on the hypothesis that energetic 
electrons emit by the gyrosynchrotron mechanism.
For this class of radio stars, characterized by a mainly dipolar magnetic field 
whose axis is tilted with respect to the rotational axis, the geometry of the 
magnetosphere and its deformation due to the stellar rotation are determined. 
The radio emitting region is determined by the physical conditions of the
magnetosphere and of the stellar wind.
Free-free absorption by the thermal plasma 
trapped in the inner magnetosphere is also considered.
Several free parameters are involved in the model, such as the size of the emitting
region, the energy spectrum and the number density of the emitting electrons, 
and the characteristics of the plasma in the inner magnetosphere.
By solving the equation of radiative transfer, along a path parallel 
to the line of sight, the radio brightness distribution and the total flux density 
as a function of stellar rotation are computed. 
As the model is applied to simulate the observed 5\,GHz lightcurves of 
\object{HD\,37479} and \object{HD\,37017}, several possible magnetosphere 
configurations are found. After simulations at other frequencies, in spite of 
the large number of parameters involved in the modeling, two solutions in the 
case of \object{HD\,37479} and only one solution in the case of \object{HD\,37017}
match the observed spectral indices. 
The results of our simulations agree with the magnetically confined 
wind-shock model in a rotating magnetosphere.
The X-ray emission from the inner magnetosphere is also computed, and found
to be consistent with the observations.
   }

   \titlerunning{3D model for radio emission from magnetic CP stars}
   \authorrunning{C. Trigilio et al.}
   \maketitle

%============================================fig 1
\begin{figure*}
\resizebox{\hsize}{!}{\includegraphics{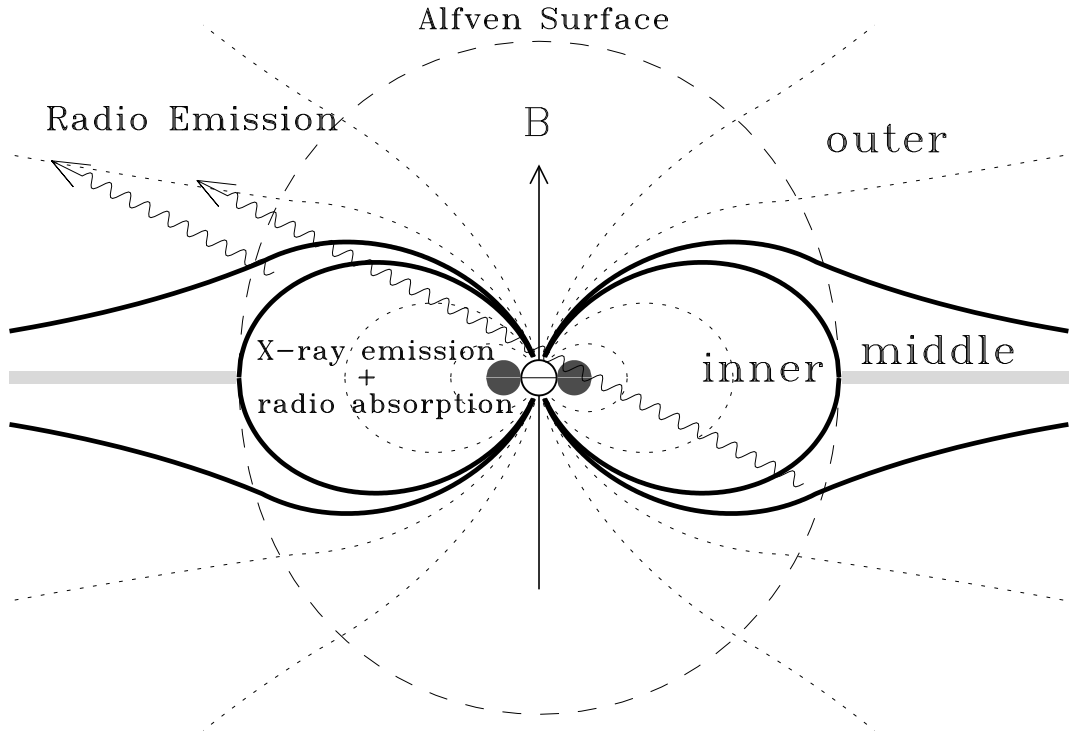}}
 \caption[ ]{
Cross-section of the magnetosphere of magnetic chemically peculiar stars: 
the stellar dipolar field (short dashed curves) is modified by the stellar 
wind; inside the Alfv\'en surface (long dashed curve), the magnetic pressure 
exceeds the kinetic pressure of the wind, and the field lines maintain a dipolar
geometry. The largest closed field line defines the ``inner" magnetosphere,
which confines the stellar wind (``dead zone"). 
Here the two stellar wind streams from opposite hemispheres of the stellar 
surface collide, leading to a shock that produces an enhancement of the
temperature of the gas and eventually to X-ray emission. 
This gas cools and accumulates in the magnetic equatorial 
plane, forming a torus-like cloud (the two filled circles close to the star). 
The open magnetic field lines just outside the inner magnetosphere
produce current sheets (shaded areas), where the electrons are accelerated
up to relativistic energies. They eventually can propagate back toward the
stellar surface following the field lines of the ``middle" magnetosphere
and emit radio radiation by the gyrosynchrotron emission process (wave arrows). 
The hot matter in the inner magnetosphere can absorb the radio radiation.
The field lines close to the magnetic poles of the star are open 
(``outer" magnetosphere), and tend to
a radial topology outside the Alfv\'en surface. Only the wind from the polar
caps can escape from the magnetosphere.
}
\label{apri_alf}
\end{figure*}

\section{Introduction}
Magnetic Chemically Peculiar (MCP) stars are commonly characterized by 
strong ($B \sim 10^3$ -- $10^4$~Gauss) and periodically-variable surface magnetic fields.
The observed magnetic variability is related to the global
magnetic field topology of these stars which is mainly
dipolar, with the magnetic axis tilted with 
respect to the rotational axis.
The observed variability is a simple consequence of stellar rotation
(Babcock \cite{Babcock49}).

These stars show evidence of anisotropic winds, 
as indicated by spectral observations of UV lines 
(Shore et al. \cite{shore_et_al}, Shore \& Brown \cite{sho_e_bro}).
Theoretical studies have demonstrated that a stellar 
wind in the presence of a dipolar magnetic field can freely
flow only from the polar regions, where the magnetic 
field lines are open (Shore \cite{shore}). Therefore in MCP stars 
the wind forms two polar jets, whereas 
at the latitudes near the magnetic equator the wind is inhibited and the matter
is trapped, forming the so called ``dead zone".
The presence of jets and circumstellar matter
can explain the emission features observed in the UV spectra and $H_\alpha$
wings (Walborn \cite{walborn}).
In some cases the mass loss rate
($\dot{M}=10^{-9}$ -- $10^{-10}~M_{\sun}\mathrm{yr}^{-1}$) and
the outflow terminal speed ($v_{\infty}\simeq 600$ km s$^{-1}$) was derived 
(Shore et al. \cite{shore_et_al}, Groote \& Hunger \cite{gro_e_hun}).

About 25\% of MCP stars also show evidence of non-thermal radio continuum 
emission (Drake et al. \cite{drake_et_al}, Linsky et al. \cite{linsky_et_al}, 
Leone et al. \cite{leone_et_al}).
It has been suggested that the physical processes responsible for 
the observed radio emission are
strictly related to the interaction between wind
and magnetic field (Linsky et al. \cite{linsky_et_al}, 
Babel \& Montmerle \cite{bab_e_mon}, hereafter BM97).
The flat spectral index 
and the observed degree of circular polarization have been interpreted 
in terms of gyrosynchrotron emission
from continuously injected mildly relativistic electrons
trapped in the stellar magnetosphere. 

The radio emission of MCP stars is variable with the same period as the 
magnetic field variability, as shown by Leone (\cite{leone}) and Leone \& Umana 
(\cite{leo_e_uma}) for \object{HD\,37479} and \object{HD\,37017}, and by Lim et al.
(\cite{lim}) for \object{HR\,5624}.
In particular, for \object{HD\,37479} and \object{HD\,37017}, 
the minimum of the radio emission coincides 
approximately with the zero of the longitudinal magnetic field, whereas the 
maxima coincide with the extremes of the magnetic field curves (Leone \& Umana 
\cite{leo_e_uma}).
The observed modulation suggests that the radio emission arises from a stable 
corotating magnetosphere. The temporal variability of the radio flux is 
probably related to the change of the orientation of the emitting region 
in the space, due to the misalignment of magnetic and rotational axes.

In this paper we present a three-dimensional gyrosynchrotron model developed 
with the aim of investigating the nature of the rotational modulation of the 
radio emission. 
This kind of study can be used to test the physical scenario proposed to 
explain the origin of radio emission from MCP stars.

\section{The model}
\label{origin}

\noindent
To explain the radio continuum and the X-ray emission from young magnetic 
B stars and MCP stars, Andr\'e et al. (\cite{andre_et_al}) proposed a model 
characterized by the interaction between the dipolar magnetic field and the 
stellar wind. 
Close to the star, where the magnetic pressure dominates over the kinetic 
pressure of the wind, the magnetic topology is mainly dipolar.
As the magnetic field strength decreases, the kinetic pressure equals the
magnetic one, defining the Alfv\'en surface, and the wind opens the magnetic field 
lines, generating a ``current sheet" in the magnetic equatorial plane
(Havnes \& Goertz \cite{hav_e_goe}).
Here the electrons of the wind can be accelerated up to relativistic energies
(Usov \& Melrose \cite{uso_mel}); eventually those electrons return to the 
star, along the field lines, thus defining the gyrosynchrotron
emitting region.
In the magnetosphere we identify three main zones:
\begin{itemize}
\item[i)] 
the ``inner" magnetosphere or ``dead zone", completely inside the Alfv\'en surface,
defined by the largest closed magnetic line; here the wind is confined;
\item[ii)]
the ``middle" magnetosphere, where the radio emission occurs, 
holding all the open magnetic field lines that generate the current sheets;
\item[iii)]
the ``outer" magnetosphere, where the wind flows freely from the magnetic
polar caps, with a magnetic topology almost radial out of the Alfv\'en surface.
\end{itemize}
\noindent
Linsky et al. (\cite{linsky_et_al}) located as possible radio emitting regions 
two tori,
one in the northern, the other in the southern magnetic hemisphere, inside the 
``middle" magnetosphere, the higher frequency being generated closer to the
star, where the magnetic field is stronger. 

The recent discovery of coherent emission at 20~cm from \object{HD\,124224} by
Trigilio et al. (\cite{trigilio}), explained in terms of
Electron Cyclotron Maser Emission (ECME), seems to confirm the hypothesis of 
relativistic electrons accelerated in the current sheets and returning
toward the star, where they are reflected back. 
In fact the ECME can occur after the magnetic mirroring when, in particular 
conditions, a loss cone anisotropy can develop (Melrose \& Dulk \cite{meldulk}).

The basic scenario of the magnetosphere of a MCP star is sketched in
Fig.~\ref{apri_alf}.
If this scenario is correct, it should  also reproduce the rotational 
modulation of the observed radio emission.
We outline the basic steps of the model in Appendix~\ref{procedure},
while the stellar reference frame and its rotation are presented in
Appendix~\ref{frame} and \ref{rotation}. The numerical computation
of the radio emission toward the Earth is presented in Appendix~\ref{emission}.
In the following, we analyze the numerical sampling of the emitting region,
and the physical conditions of the magnetosphere.

\subsection{Sampling}
The space surrounding the star is sampled in 
a 3-D cubic grid (see Appendix~\ref{frame}) and all the
physical quantities relevant for the numerical computation of the radio emission
(such as magnetic field, electron density,...)
are evaluated in each grid point. 

Since the frequency of the gyrosynchrotron radiation is
proportional to the gyration frequency of the electrons, which in turn is
proportional to the magnetic field intensity,
different frequencies are mainly emitted in 
different regions of the magnetosphere at different distances from the stellar 
surface, because of the strong gradient of $B$ in a dipole.
Therefore the modelled region is a cube whose overall size is a function of the 
frequency. This cube is then sampled into smaller cubes, whose size is
chosen so that any physical parameter inside them can be assumed constant.

The reliability of the model is obviously greater if the 
size of the cube element is smaller
but, on the other hand, the computational time increases strongly. 
The compromise between
sampling and computational time is chosen on the basis of the convergence
of the model (the flux for instance) as a function of the sampling
toward an asymptotic value.
In particular, to model the 5~GHz radiation from the MCP stars,
the overall linear size of the grid is $25-30$ stellar radii, divided into about
$80-100$ cube elements. 

\subsection{Location of the Alfv\'en surface}
\label{loc_alf}
%============================================fig 2
\begin{figure}
\resizebox{\hsize}{!}{\includegraphics{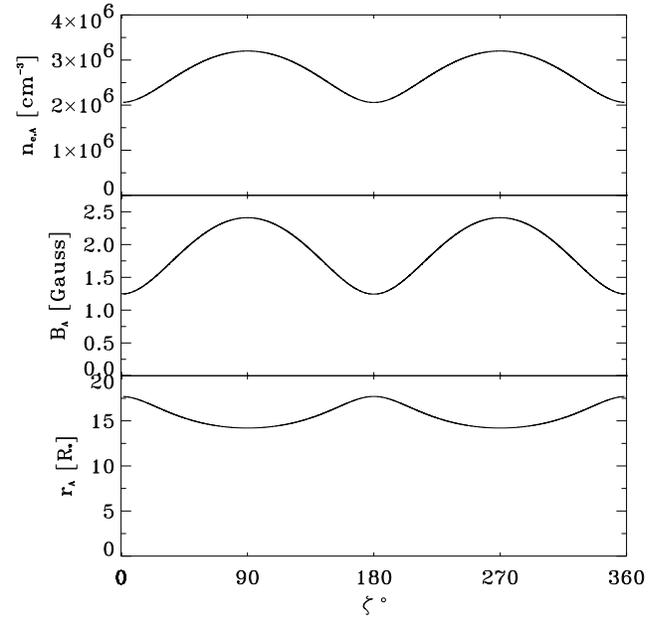}}
 \caption[ ]{Thermal electron number density (top panel),
 magnetic field strength 
(medium panel) and equatorial radius of the Alfv\'en surface (low panel) versus
magnetic longitude $\zeta$. 
They are derived with 
$R_{\ast}=4~R_{\sun}$, $P_\mathrm{rot}=1$~day,  
$B_\mathrm{p}=10^4$ Gauss, $\beta=60\degr$,
$\dot{M}= 10^{-9}~M_{\sun}\mathrm{yr}^{-1}$ and 
$v_{\infty}=600$~km s$^{-1}$, see Sect.~\ref{loc_alf} for details.
}
\label{nbr_alf}
\end{figure}
%=======================================================

\noindent
The equatorial radius of the Alfv\'en surface ($r_\mathrm{A}$) may be
estimated by equating the kinetic energy density $\varepsilon_\mathrm{w}$
and the magnetic energy density $\frac {B^2}{8\pi}$
(Usov \& Melrose \cite{uso_mel}):
\begin{equation}
\varepsilon_\mathrm{w} =\frac {B^2}{8\pi} 
\label{eq_1}
\end{equation}
The term $\varepsilon_\mathrm{w}$ includes both radial and rotational
components of the kinetic energy of the wind: 
\begin{equation}
\varepsilon_\mathrm{w} = \frac{1}{2}\rho v^2 + \frac{1}{2}\rho \omega ^2 d^2
\label{dens_wind}
\end{equation}
where $\rho$ and $v$ are the density and the speed of the wind respectively, 
$\omega$ is the angular velocity of the star and $d$ the distance of the 
generic point of the magnetic equator from the rotational axis,
given by:
\begin{displaymath}
d=r \sqrt{1-\sin ^2 \beta \cos ^2 \zeta}
\end{displaymath}
where $\beta$ is the obliquity of the magnetic field, 
$\zeta$ is the magnetic longitude
(defined to be zero in the
line located by rotational and magnetic equator planes)
and $r$ the radial distance from the centre of the star.
The gas density $\rho$ in the ``outer" region  can be estimated by the 
continuity equation:
\begin{equation}
\rho = \frac{\dot{M}}{4\pi r^2 v(r)}
\label{ro}
\end{equation}
\noindent
where $\dot{M}$ is the mass loss rate. The wind speed $v$ is a function of
the distance from the star. 
For the sake of simplicity, we use the wind speed function derived by
Castor \& Simon (\cite{cas_e_sim}):
\begin{equation}
v(r) =v_{\infty} \left(1 - \frac{R_{\ast}}{r} \right)
\label{v}
\end{equation}
\noindent
This equation is similar to the velocity law
$v(r)=v_{\infty}\left(1 - R_{\ast}/r \right)^{0.8}$
theoretically predicted by Friend \& Abbott (\cite{fri_e_abb}) for a 
radiatively driven stellar wind.
However, we emphasize that the above dependence of the stellar
wind on the distance applies only for the ``outer" and ``middle" magnetosphere,
while we expect a different velocity and density distribution
in the ``dead zone" (see Sect.~\ref{term_eff}), as the close magnetic flux 
tubes strongly influence the dynamics of the trapped wind (BM97).

To estimate the magnetic energy density in the equatorial plane
of the ``inner" magnetosphere, we use the 
expression of the strength for a dipolar field:
\begin{equation}
B =\frac{1}{2}B_\mathrm{p} \left(\frac{R_{\ast}}{r}\right)^3
\label{campo_m}
\end{equation}
where  $B_\mathrm{p}$ is the strength of the magnetic field at the pole of
the star.

Substituting Eqs.(\ref{dens_wind}) and (\ref{campo_m}) 
in Eq.(\ref{eq_1}) we derive the equatorial 
Alfv\'en radius $r_\mathrm{A}$ as a function of the magnetic longitude $\zeta$
searching the real roots.
The equatorial Alfv\'en radius ($r_\mathrm{A}$), the corresponding magnetic 
field strength ($B_\mathrm{A}$) and the thermal electron number density
($n_\mathrm{e,A}$)
are plotted in Fig.~\ref{nbr_alf} as a function of $\zeta$. 
They are derived for
$R_{\ast}=4~R_{\sun}$, rotational period $P_\mathrm{rot}=1$~day, polar field 
$B_\mathrm{p}=10^4$ Gauss, obliquity $\beta=60\degr$,
mass loss $\dot{M}= 10^{-9}~M_{\sun}$ yr$^{-1}$ and 
$v_{\infty}=600$~km s$^{-1}$ (Drake et al. \cite{drake_et_al}).
Figure \ref{nbr_alf} shows how the stellar rotation 
significantly deforms the magnetosphere.

%============================================fig 3
\begin{figure}
\resizebox{\hsize}{!}{\includegraphics{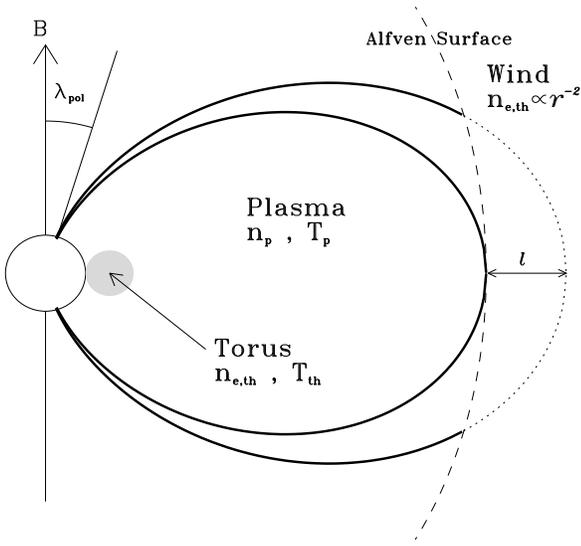}}
 \caption[ ]{Section of the magnetic shell which delimits the emitting
magnetospheric volume. The parameter $l$ determines
the outer magnetic field line, 
and, consequently, the size of the radio emitting region. 
In the outer magnetosphere (wind zone) the density decreases with
distance.
Close to the star, a cooler absorbing material is present, which we ascribe
to a torus of constant density $n_\mathrm{e,th}$ and temperature $T_\mathrm{th}$.
The plasma temperature and density in the inner magnetosphere 
($T_\mathrm{p}$ and $n_\mathrm{p}$) are discussed
in Sect.~\ref{term_eff}. 
The parameter $l$ also defines also the width $\lambda_\mathrm{pol}$ of the polar
caps, from where the wind can flow freely.
}
\label{sezione}
\end{figure}
%=======================================================
\subsection{
Energetic electron distribution and location of the ``middle" magnetosphere}

\noindent
The present model assumes that electrons are accelerated 
in equatorial current sheets, drawn in Fig.~\ref{apri_alf}
as shaded areas outside the Alfv\'en surface.
The volume of the magnetosphere where the energetic electrons can propagate 
is defined in Fig.~\ref{sezione} by thick solid lines. 

The model assumes that, in each point of the ``middle" magnetosphere, 
the emitting electrons are isotropically
distributed in pitch angle and have a power law energy distribution
\begin{equation}
N(\gamma) \propto N_\mathrm{e} (\gamma-1)^{-\delta}
\label{power_law}
\end{equation} 
where $\gamma$ is the Lorentz factor and $N_\mathrm{e}$ is the total number
density of the energetic electrons.
As low energy cutoff, we choose $\gamma=1.2$, corresponding to 100~keV.
This choice does not influence the computation of the gyrosynchrotron emission, 
as discussed by Klein (\cite{klein}). We choose the exponent $\delta$ 
considering that in the solar flares, where an acceleration process
as in MCP stars is supposed, microwave and hard X-ray emission are
assumed to be generated by the same population of non-thermal electrons. 
The spectrum of the hard X-ray photons is proportional to 
$\nu^{-\left(\delta+\frac{1}{2}\right)}$ 
when the electron population is described by Eq.(\ref{power_law}). Since the 
observed hard X-ray spectrum in solar flares
is a power law with an exponent ranging from 
$-3$ to $-4$ (Longair \cite{longair}),
$\delta$ should be in the range $2.5$ --$\,3.5$. At the maximum of the solar
flares, i.e. close to the impulsive event, the spectrum is generally quite flat
($\delta \approx 2$ --$\,3$), and eventually it becomes steeper, as the more
energetic electrons lose their energy faster.
If in MCP stars the electron acceleration occurs in the current sheets 
outside the Alfv\'en radius, then the non-thermal electrons injected toward the 
star do not lose much energy via microwave emission, as they quickly cross
the emission region (tens of seconds).
Therefore they will maintain the original flat energetic spectrum.
In the following, we will consider a spectrum distribution index 
$\delta$ in the range $2$ --$\,4$.

The electron number density remains almost constant as the electrons
propagate inside the magnetosphere, as
a consequence of two combined effects (Hasegawa \& Sato \cite{has_e_sat}): 1)
particle losses due to magnetic mirroring and 2) compression due to the 
decreasing volume of the magnetic shell.

We assume that the number density $N_\mathrm{e}$ is a fraction of the
thermal plasma density $n_\mathrm{e,A}$ at the Alfv\'en surface. This ratio is a
free parameter in our model, since it depends on the efficiency of the
acceleration process, and we assume $N_\mathrm{e}/n_\mathrm{e,A}$ in the
range $10^{-4}$ --$\,1$.

The ``middle" magnetosphere is defined by two magnetic lines (Fig.~\ref{sezione}): 
the first is the
largest closed line of the ``inner" magnetosphere, tangent to the Alfv\'en 
surface in the plane of the magnetic equator;
the latter is defined by the size of the current sheet, 
which depends on the particular physical conditions
of the magnetosphere, and which we consider a free parameter of our model.

In a magnetic dipole, the equation of a field line is given by
\begin{displaymath}
r = L \cos ^2 \lambda
\end{displaymath}
where $L$ is the distance between the center of the star
and the point where the field line crosses the magnetic 
equatorial plane; $\lambda$ is the angle between this
plane and the radius vector $r$.
The middle magnetosphere is therefore identified by the points
belonging to magnetic field lines having an
equatorial distance $L$
between $r_\mathrm{A}$ and $r_\mathrm{A} + {l}$, where ${l}$
is a free parameter in our calculations.

%============================================fig 4
\begin{figure}
\resizebox{\hsize}{!}{\includegraphics{tr.f4a}}
\resizebox{\hsize}{!}{\includegraphics{tr.f4b}}
 \caption[ ]{ Effective magnetic field variations (Bohlender \cite{bohl_87})
and 5 GHz lightcurves of \object{HD\,37479} and \object{HD\,37017}:
($\times$) data from Drake et al. (\cite{drake_et_al}), ($\bullet$) Leone \& 
Umana (\cite{leo_e_uma}), ($\diamond$) this paper.
}
\label{2cp}
\end{figure}
%=======================================================

\subsection{Thermal plasma in the inner magnetosphere}
\label{term_eff}

\noindent
The presence of thermal plasma trapped in the dead zone may affect the stellar
radio emission because of the free-free absorption
(Andr\'e et al. \cite{andre_et_al}).
If its density is high enough, this plasma
can partially absorb the radiation and this could result in an additional
modulation of the radio lightcurve, since this effect is greater for a
particular orientation of the magnetosphere.
Moreover, if the density of the plasma is higher close to the star,
the highest frequencies, which are emitted in the inner part of the magnetosphere,
will be strongly absorbed.
In contrast, the lowest frequencies, generated far from the star, 
are less affected by the plasma absorption.
Therefore, the density distribution inside the inner magnetosphere
can affect either the modulation of the radio lightcurve, or the
spectrum of the source.
In particular, it is important to define the extent of the absorbing
matter into the inner magnetosphere, the density distribution and the
temperature.

Effects of the trapped plasma have been studied at other wavelengths,
as it causes also the absorption of resonance lines and of continuum
when the star is seen with the equator edge-on.
Smith \& Groote (\cite{smith2001}), from the UV spectra of several 
magnetic early-B stars, found some 
absorbing material, which they interpreted as a ``cloud" along the line of sight,
characterized by a column density ($n_\mathrm{col}$) in the range 
$10^{22 \,-\,23}\mathrm{\,cm}^{-2}$, temperature in the range
$15\,000$ --$\,31\,000\,\mathrm{\,K}$ and average density $10^{12 \,-\,13} \mathrm{\,cm}^{-3}$.

With the aim of explaining the X-ray emission from \object{IQ~Aur},
BM97 developed a magnetically confined
wind-shock model, where the radiatively-driven winds from the two magnetic
hemispheres collide, producing a shock that leads to an enhancement of the
temperature up to $10^{6 \,- \,7}$ K in the post-shock region, where X-ray radiation
is emitted. 
They also predicted, from their X-ray model, the existence of a ``disk" 
along the equator, made of post-shock cooling material, which could in 
particular explain the $15.4^{d}$ rotational modulation of
\object{$\theta^1$~Ori~C} seen by {\em ROSAT} (Babel \& Montmerle \cite{BM2}).
In this model, the cloud observed by Smith \& Groote (\cite{smith2001})
could be the inner part of the equatorial disk.
Also Donati et al. (\cite{donati2001}) interpreted the X-ray emission and the
variability of \object{$\beta$~Cep} in the framework of this model.
The physical scenario of such a magnetosphere has been summarized
by Montmerle (\cite{Mont}) (see also Fig.~\ref{apri_alf}).

Once a steady state of this process has been reached,
the dynamics of the wind and the physical characteristics of the matter
along lines of force of the magnetic field are ruled by the equilibrium
of the total pressure.
This implies that
\begin{equation}
n_\mathrm{p}k_\mathrm{B}T_\mathrm{p}=p_\mathrm{ram}
\label{equil_press}
\end{equation}
with $n_\mathrm{p}$ and
$T_\mathrm{p}$ respectively the number density and the temperature 
of the plasma in the post-shock region 
and $p_\mathrm{ram}=\rho v^2$ is the ram pressure of the wind.
The density $\rho$ and the speed $v$ come from Eqs.(\ref{ro}) and (\ref{v}).
If $\dot{M}=10^{-9}~M_{\sun}\mathrm{yr}^{-1}$ and $v_{\infty}=600$ km s$^{-1}$,
then $p_\mathrm{ram}\approx 10^{-1}$ --$\,10^{-2}\mathrm{\,g\,cm^{-1}s^{-2}}$.
Assuming that the radio absorbing region is the post-shock
region of BM97, the previous relations give
$n_\mathrm{p}T_\mathrm{p}\approx 10^{14}$ (in c.g.s.); if 
$T_\mathrm{p}\approx 10^6$ K, $n_\mathrm{p}\approx 10^8\mathrm{\,cm}^{-3}$.

If, in addition, the post-shock region were also responsible for the
absorption of the UV lines and continuum (Smith \& Groote, \cite{smith2001}),
a column density of $10^{22}\mathrm{\,cm}^{-2}$ should lead to a characteristic
size of the post-shock region given by
$n_\mathrm{col}/n_\mathrm{p}$
that, considering the Eq.(\ref{equil_press}), is 
$\approx 10^{22}/n_\mathrm{p}\approx 10^{8}T_\mathrm{p}$.
Therefore, to be responsible also for the absorption of both UV features and 
radio radiation, the X-ray emitting region (with $T_\mathrm{p}\approx 10^6$ K)
should be very large ($>10^{14}\mathrm{\,cm}$), at least
some hundreds times the stellar radius or some tens the Alfv\'en radius.
In few words, so high a column density cannot be reached inside the
Alfv\'en radius unless the density is very high.

The other possibility is that the X-ray emitting region (BM97)
is not responsible for the UV absorptions, and the
only valid relation between $T_\mathrm{p}$ and $n_\mathrm{p}$ is 
Eq.(\ref{equil_press}). If $T_\mathrm{p}\approx 10^{6 \,-\,7}$ K, we find
$n_\mathrm{p}\approx 10^{8 \,-\,7}\mathrm{\,cm^{-3}}$.
Briefly, starting from the stellar surface, we have:
\begin{itemize}
\item[a)] a dense region ($n_\mathrm{e,th}\approx 10^{12}\mathrm{\,cm}^{-3}$)
at stellar temperature, ($T_\mathrm{th}\approx T_\mathrm{eff}$)
with a size of about $10^{11}$ cm, less than the stellar radius:
this region is responsible for the UV absorption;
for simplicity, a torus shape has been assumed to account
for the two equatorial matter distributions, with density $n_\mathrm{e,th}$ and
temperature $T_\mathrm{th}$ assumed constant;
\item[b)] a less dense region ($n_\mathrm{p}\approx 10^{7 \,-\,8}\mathrm{\,cm}^{-3}$)
at high temperature ($T_\mathrm{p}\approx 10^{6 \,-\,7}$ K), whose maximum extent
is defined by the magnetic confinement
($B^2/8\pi \approx n_\mathrm{p}kT_\mathrm{p}$).
Since the magnetic field is given by Eq.(\ref{campo_m}),
the maximum size is given by
$$
L=\left(\frac{B_\mathrm{p}^2}
{16\pi n_\mathrm{p}k_\mathrm{B}T_\mathrm{p}}\right)^{1/6}
$$
which gives $L=23$ --$\,18\,R_{\ast}$ respectively for
$B_\mathrm{p}=10\,000$ --$\,5\,000$ Gauss. In any case $L$ is greater than the
Alfv\'en radius (Fig.~\ref{nbr_alf}, lower panel). We can therefore assume that
this region fills all the inner magnetosphere.
\end{itemize}
Figure \ref{sezione} also outlines this scenario.
The actual situation can, however, be a little different 
since the above picture is valid
in the hypothesis of a non-rotating star. 
In fact, BM97 showed that the temperature of the plasma in the 
inner magnetosphere remains almost
constant if the star does not rotate, but can increase linearly outward in the
case of a fast rotation (with the dipole axis aligned with the rotational one, i.e
with $\beta=0$).
Since the plasma is in pressure equilibrium
with the wind ram pressure, we have to add to the first term of
Eq.(\ref{equil_press}) a kinetic term that accounts for the rotational energy,
in a similar way as in Eq.(\ref{dens_wind}). In any case, if the rotation
is considered, the density should decrease outward.
In the case that
$\beta \neq 0$ the situation is more complex, but when $\beta \approx 90\degr$
the rotational axis lies close to the magnetic equatorial plane,
and the regions 
close to the rotational axis can be considered to be not rotating, while
other regions can be rotating fast; the result is intermediate between a
non-rotating and a rotating magnetosphere.

The presence of thermal plasma has been considered in the calculation of the
emitted radiation (see Appendix~\ref{emission}), including free-free thermal
emission and absorption (Dulk \cite{dulk}) as well as the Razin effect.

%___________________________________tabella
\begin{table}
\caption{Stellar parameters}
\label{tab_cp}
\begin{tabular}[]{lccccc}
\hline
\hline
\multicolumn{6}{l}{Measured parameters}\\
\hline
%~&~&~&~&~&~&~&\\
~	&$D$	&$P_\mathrm{rot}$	&$B_\mathrm{p}$	&$i\degr$&$\beta\degr$
	\\ %~&~&~&~&~&~&~&\\
~&{\footnotesize [pc]} &{\footnotesize[day]} &{\footnotesize
[Gauss]} &~&~	\\
%~&~&~&~&~&~&\\
\hline
&~&~&~&~&\\
{\footnotesize HD\,37479 }	&{\footnotesize $352^{\dag}$}	&{\footnotesize
$1.19081^{a}$}	&{\footnotesize$6800^{a}$}	&{\footnotesize $72^{b}$}		&{\footnotesize $56^{b}$} \\
{\footnotesize HD\,37017 }	&{\footnotesize $373^{\dag}$}	&{\footnotesize 
$0.901195^{a}$}	&{\footnotesize$7700^{a}$}	&{\footnotesize$25^{b}$}		&{\footnotesize $65^{b}$}	\\
&~&~&~&~&\\
\hline
\multicolumn{6}{l}{Derived parameters}\\
\hline
~	&\multicolumn{2}{c}{$r_\mathrm{A}$ (min--max)} &\multicolumn{2}{c}
{$n_\mathrm{e,A}$ (min--max)}\\
~	&\multicolumn{2}{c}{[$R_{\ast}$]} &\multicolumn{2}{c}{[$\times 10^6$ cm$^{-3}$]}\\
\hline
&~&~&~&~&\\
{\footnotesize HD\,37479 }
	&\multicolumn{2}{c}{\footnotesize 13.18--15.75}
	&\multicolumn{2}{c}{\footnotesize 2.6--3.7 }	\\
{\footnotesize HD\,37017 }
	&\multicolumn{2}{c}{\footnotesize 12.59--16.36}
	&\multicolumn{2}{c}{\footnotesize 2.4--4.1}	\\	

~&~&~&~&\\
\hline
\end{tabular}
\begin{list}{}{}
\item[$^{\dag}$] Hipparcos main catalogue
\item[$^a$] Bohlender et al. (\cite{bohl_87})
\item[$^b$] Shore \& Brown (\cite{sho_e_bro})
\end{list}

\end{table}
%___________________________________________

\section{Application to \object{HD\,37479} and \object{HD\,37017}}
\label{appli}

\noindent
The model developed in this paper has been used
to simulate the radio lightcurves of two MCP stars:
\object{HD\,37479} (=$\sigma$~Ori~E) and \object{HD\,37017}.
%%============================================fig 5
\begin{figure*}
\resizebox{\hsize}{!}{\includegraphics{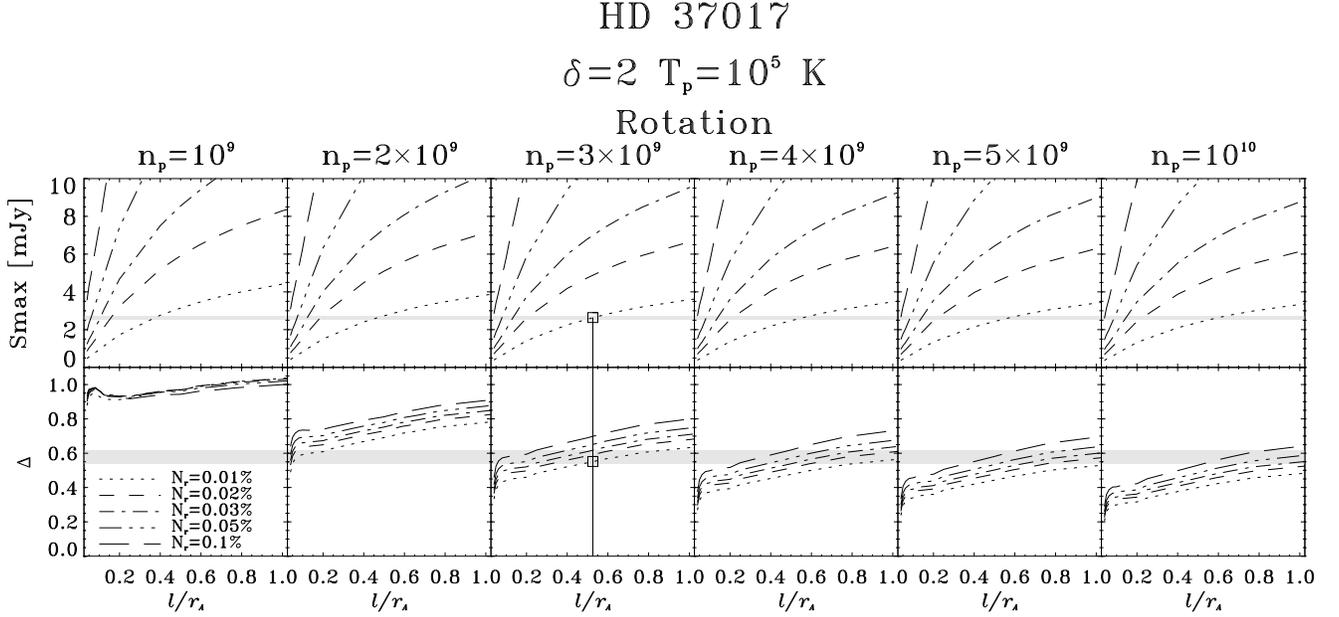}}
 \caption[ ]{HD 37017: 
examples of search of solutions. Light curve characteristics:
maximum flux density (top panels) and ratio $\Delta$ (lower panels) as a function
of the magnetic shell thickness $l$ normalized to the stellar Alfv\`en radius
$r_A$. 
In these examples, the spectral index of the relativistic electrons ($\delta$) 
has been fixed to 2, and the temperature of the plasma at the base of the inner 
magnetosphere ($T_\mathrm{p}$) to $10^{5}$\,K. The rotation of the star has been
considered. From left to right: solutions for increasing values of the
plasma density ($n_\mathrm{p}$) at the base of the inner magnetosphere.
In each pair of panels, $S_\mathrm{max}$ and $\Delta$ are computed 
for different number density of the emitting electrons ($N_\mathrm{r}$)
as percentage of the number density of thermal plasma at the Alfv\'en point
$n_\mathrm{e,A}$ (different line styles).
A valid solution is given if the computed $S_\mathrm{max}$ and $\Delta$
are within the observed values (shaded areas) for the same value of
$l/r_\mathrm{A}$. 
The good solution is indicated by a vertical line and two squares.}
\label{smz_017}
\end{figure*}
%%=======================================================

The 5 GHz emission of these two stars appears to vary with the rotational 
period.
Fig.~\ref{2cp} shows the effective magnetic field $B_\mathrm{e}$
and the observed 5 GHz flux densities $S_5$ as reported by Leone \& Umana
(\cite{leo_e_uma}).
We add to these data the new results of two VLA observations: 
\object{HD\,37479} was observed on 29 April 1995 (21:29--21:47 UT)
and \object{HD\,37017} on 29 May 1995 (18:42--19:06 UT). We reduced and
analysed the data following the procedures used by Leone \& Umana
(\cite{leo_e_uma}). We measured as flux densities
respectively $3.0\pm 0.1$ mJy at phase $\Phi=0.545$ and $1.9\pm 0.2$ mJy at
$\Phi=0.16$, with rotational phases computed by using the ephemeris determined 
by Bohlender et al. (\cite{bohl_87}). Those data are reported in Fig.~\ref{2cp}
as diamonds symbols.
Leone \& Umana (\cite{leo_e_uma}) noted that,
while the radio data of \object{HD\,37017}
can be fitted by a single sinusoidal like curve, for \object{HD\,37479}
the radio variation can have a double-wave behaviour, as for the variations
of the helium abundance in those stars. They argued that the radio
emission is probably related to $|B_\mathrm{e}|$,
with the minimum corresponding approximately with the zero of the longitudinal
magnetic field. 

Regarding the X-ray emission, that could arise from the inner magnetosphere,
both stars have been observed in the {\em ROSAT} all-sky survey
(Bergh\"ofer et al. \cite{RASS}).
While \object{HD\,37017} is reported as a non-detection
($\lg L_\mathrm{X}<30.7$), 
\object{HD\,37479} is a doubtful detection (Drake et al. \cite{drake1994}) since
it is a visual binary with a hotter primary component closer than $1\arcmin$,
likely the predominant contributor to the observed X-ray emission.

To simulate the 5 GHz radio emission of these two MCP
stars, we fix a stellar radius of $R_{\ast}=4~R_{\sun}$
(Shore \& Brown \cite{sho_e_bro}) and an effective temperature
$T_\mathrm{eff}=24\,000$~K (Leone \cite{leone}). 
The other physical stellar parameters are reported in Table~\ref{tab_cp}.
We set the mass loss rate and the terminal wind speed  
respectively to $\dot{M}=10^{-9}~M_{\sun}\mathrm{yr}^{-1}$ and
$v_{\infty}=600$ km s$^{-1}$, according to Drake et al. (\cite{drake_et_al}).
The above mass loss rates are not the actual values, but
refer to the case of a spherical wind. In fact, the wind is confined by
the magnetosphere, leading to the formation of the torus as discussed in
Sect.~\ref{term_eff}, and it can escape only from the two polar
caps defined by the largest closed field line (see Sect.~\ref{loc_alf} and
Fig.~\ref{sezione}).
Using these observing constraints we can univocally define the equatorial
radius of the Alfv\'en surface and the local thermal plasma
densities. The maximum and minimum equatorial Alfv\'en radius ($r_\mathrm{A}$)
and the corresponding thermal
electron number density ($n_\mathrm{e,A}$) for \object{HD\,37479} and
\object{HD\,37017} are also reported in Table~\ref{tab_cp}.
The actual mass loss rate $\dot{M}_\mathrm{act}$ is computed
considering that the polar caps have an aperture $\lambda_\mathrm{pol}$
defined by the intersection between the stellar surface and the magnetic
field line that identifies the ``outer" magnetosphere (see Fig.~\ref{sezione}).
Using the above stellar 
wind parameters, we get $\lambda_\mathrm{pol}\approx 15\degr$, 
so that the area of the two caps is only $2\%$ of the total
surface, giving $\dot{M}_\mathrm{act}\approx 10^{-11}M_{\sun}\mathrm{yr}^{-1}$.

The aim of our analysis is to derive possible configurations of the magnetosphere,
represented by the free parameters of the model 
($l$, $N_\mathrm{e}$, $\delta$, $T_\mathrm{p}$ and $n_\mathrm{p}$),
which result in simulated lightcurves at 5 GHz able to reproduce
the observations. The search of these combinations of the free 
parameters, computing the radio flux at different values of the
rotational phase,
could be very time-consuming so, to restrict the computing time,
we simulate the flux density of the two stars
only for particular stellar orientations, coinciding
with the extremes and the nulls of the magnetic field curve.
The characteristic of the observed radio lightcurves may be
summarized by the maximum flux density $S_\mathrm{max}$ and the ratio
$\Delta=S_\mathrm{min}/S_\mathrm{max}$, that expresses the depth of the
modulation, being minimum for high modulation.

The lightcurve characteristics derived by fitting the observing
data for \object{HD\,37479} and \object{HD\,37017} are, respectively:\\
$S_\mathrm{max}= 4.0 \pm 0.2$ mJy, $\Delta =0.65 \pm 0.06$ and \\
$S_\mathrm{max}= 2.6 \pm 0.1$ mJy, $\Delta =0.58 \pm 0.04$.

\begin{table*}
\caption{With and without rotational effects, the 5 GHz radio lightcurves of 
HD\,37479 and HD\,37017
are reproduced for the listed values of:
spectral index of non-thermal electron energy distribution ($\delta$); 
temperature ($T_\mathrm{p}$) and
thermal electron density ($n_\mathrm{p}$) of the plasma in the post-shock region 
as extrapolated to the stellar
surface; equatorial thickness in Alfv\'en radius units ($l/r_\mathrm{A}$); and 
total number density of non-thermal
electrons ($N_\mathrm{r}$). The asterisk in the last column indicates the possible
solutions that reproduce the almost flat observed spectral index.}
\label{solutions}
\begin{tabular}[]{l|cr|cr|rc|rc}
\hline
\hline
 Star &  Rotation & $\delta$ & $T_\mathrm{p}$[K] & $n_\mathrm{p}$[cm$^{-3}$] & $l/r_\mathrm{A}$ & $N_\mathrm{r}$[cm$^{-3}$] & Spectral index & \\
\hline
HD\,37479 & No  &  2 & $1.0\times 10^{ 5}$  & $>1.0\times 10^{ 9}$  & 0.74 & $6.3\times 10^{ 2}$  & --0.40 -- 0.10 & \\
%          & No  &  2 & $1.0\times 10^{ 5}$  & $1.0\times 10^{10}$  & 0.74 & $6.3\times 10^{ 2}$  & $-0.41 -- \,0.09$ & \\
          & Yes &  2 & $1.0\times 10^{ 5}$  & $1.0\times 10^{ 9}$  &  0.05 & $3.1\times 10^{ 3}$  & --0.18 -- 0.31 & \\
          & Yes &  2 & $1.0\times 10^{ 5}$  & $2.0\times 10^{ 9}$  &  0.40 & $6.3\times 10^{ 2}$  & --0.10 -- 0.34 & $\ast$ \\
          & Yes &  2 & $1.0\times 10^{ 5}$  & $3.0\times 10^{ 9}$  &  0.83 & $4.5\times 10^{ 2}$  & --0.07 -- 0.21 & $\ast$ \\
          & Yes &  3 & $1.0\times 10^{ 5}$  & $1.0\times 10^{ 9}$  &  0.11 & $3.1\times 10^{ 4}$  & --0.07 -- 0.61 & \\
          & Yes &  3 & $1.0\times 10^{ 5}$  & $2.0\times 10^{ 9}$  &  0.77 & $9.4\times 10^{ 3}$  & --0.04 -- 0.56 & \\
          & Yes &  4 & $1.0\times 10^{ 5}$  & $1.0\times 10^{ 9}$  &  0.27 & $3.1\times 10^{ 5}$  &  0.05 -- 0.74 & \\
          & Yes &  4 & $1.0\times 10^{ 5}$  & $2.0\times 10^{ 9}$  &  1.05 & $1.5\times 10^{ 5}$  &  0.11 -- 0.69 & \\
	  &&&&&&&&\\
          & No  &  2 & $1.0\times 10^{ 6}$  & $5.0\times 10^{ 8}$  &  0.36 & $9.4\times 10^{ 2}$  &  0.12 -- 0.75 & \\
          & No  &  2 & $1.0\times 10^{ 6}$  & $1.0\times 10^{ 9}$  &  0.74 & $6.3\times 10^{ 2}$  &  0.09 -- 0.54 & \\
          & No  &  3 & $1.0\times 10^{ 6}$  & $3.0\times 10^{ 8}$  &  0.18 & $3.1\times 10^{ 4}$  &  0.06 -- 0.87 & \\
          & No  &  3 & $1.0\times 10^{ 6}$  & $4.0\times 10^{ 8}$  &  0.56 & $1.5\times 10^{ 4}$  &  0.09 -- 0.92 & \\
          & No  &  3 & $1.0\times 10^{ 6}$  & $5.0\times 10^{ 8}$  &  0.91 & $1.2\times 10^{ 4}$  &  0.12 -- 0.99 & \\
          & No  &  4 & $1.0\times 10^{ 6}$  & $1.0\times 10^{ 8}$  &  0.12 & $7.8\times 10^{ 5}$  &  0.02 -- 0.81 & \\
          & No  &  4 & $1.0\times 10^{ 6}$  & $2.0\times 10^{ 8}$  &  0.12 & $7.8\times 10^{ 5}$  &  0.11 -- 0.93 & \\
          & No  &  4 & $1.0\times 10^{ 6}$  & $3.0\times 10^{ 8}$  &  0.45 & $3.1\times 10^{ 5}$  &  0.16 -- 0.96 & \\
          & No  &  4 & $1.0\times 10^{ 6}$  & $4.0\times 10^{ 8}$  &  0.83 & $2.2\times 10^{ 5}$  &  0.21 -- 1.05 & \\
          & Yes &  4 & $1.0\times 10^{ 6}$  & $1.0\times 10^{ 8}$  &  0.05 & $1.5\times 10^{ 6}$  & --0.09 -- 0.58 & \\
          & Yes &  4 & $1.0\times 10^{ 6}$  & $1.0\times 10^{ 9}$  &  0.05 & $1.5\times 10^{ 6}$  & --0.09 -- 0.60 & \\
	  &&&&&&&&\\
          & No  &  4 & $1.0\times 10^{ 7}$  & $1.0\times 10^{ 7}$  &  0.07 & $1.5\times 10^{ 6}$  & --0.04 -- 0.71 & \\
          & No  &  4 & $1.0\times 10^{ 7}$  & $1.0\times 10^{ 8}$  &  0.07 & $1.5\times 10^{ 6}$  & --0.04 -- 0.71 & \\
          & Yes &  4 & $1.0\times 10^{ 7}$  & $>1.0\times 10^{ 7}$  &  0.09 & $7.8\times 10^{ 5}$  & --0.11 -- 0.66 & \\
%          & Yes &  4 & $1.0\times 10^{ 7}$  & $1.0\times 10^{ 8}$  &  0.09 & $7.8\times 10^{ 5}$  & $-0.11 -- \,0.66$ & \\%
\hline
HD\,37017 & No  &  2 & $1.0\times 10^{ 5}$  & $>1.0\times 10^{ 9}$  &  0.59 & $4.5\times 10^{ 2}$  & --0.44 -- 0.09 & \\
%          & No  &  2 & $1.0\times 10^{ 5}$  & $1.0\times 10^{10}$  &  0.59 & $4.5\times 10^{ 2}$  & $-0.45 -- \,0.09$ & \\
          & Yes &  2 & $1.0\times 10^{ 5}$  & $3.0\times 10^{ 9}$  &  0.53 & $3.1\times 10^{ 2}$  & 0.00 -- 0.34 & $\ast$ \\
          & Yes &  3 & $1.0\times 10^{ 5}$  & $2.0\times 10^{ 9}$  &  0.62 & $4.7\times 10^{ 3}$  & 0.14 -- 0.68 & \\
          & Yes &  4 & $1.0\times 10^{ 5}$  & $2.0\times 10^{ 9}$  &  0.67 & $9.0\times 10^{ 4}$  & 0.08 -- 0.86 & \\
	  &&&&&&&&\\
          & No  &  2 & $1.0\times 10^{ 6}$  & $5.0\times 10^{ 8}$  &  0.05 & $3.1\times 10^{ 3}$  &  0.27 -- 1.10 & \\
          & No  &  2 & $1.0\times 10^{ 6}$  & $1.0\times 10^{ 9}$  &  0.59 & $4.5\times 10^{ 2}$  &  0.22 -- 0.60 & \\
          & No  &  3 & $1.0\times 10^{ 6}$  & $4.0\times 10^{ 8}$  &  0.40 & $9.4\times 10^{ 3}$  &  0.29 -- 1.18 & \\
          & No  &  4 & $1.0\times 10^{ 6}$  & $3.0\times 10^{ 8}$  &  0.36 & $1.5\times 10^{ 5}$  &  0.29 -- 1.12 & \\

\hline
\end{tabular}
\end{table*}

The simulations for the two stars are performed by using different
combinations of the free parameters of the model:
\begin{itemize}
\item $l$: equatorial thickness of the magnetic shell (Fig.~\ref{sezione}),
such that $l/r_\mathrm{A}=0.025$ -- 1;
\\
\item $N_\mathrm{e}$: total number density of the non-thermal
electrons, in the range $10^2$ --$\,10^6$cm$^{-3}$, 
with $N_\mathrm{e}<n_\mathrm{e,A}$;
\\
\item $\delta$: spectral index of the non-thermal electron energy
distribution (Eq.(\ref{power_law})), in the range 2 -- 4;\\
\item $T_\mathrm{p}$ and $n_\mathrm{p}$: temperature and number density
of the plasma in the inner magnetosphere (the post-shock region of BM97); 
they are chosen so that
$T_\mathrm{p}n_\mathrm{p} = p_\mathrm{ram}/k_\mathrm{B}$ is in the range
$10^{14}$ -- $10^{15}$ (in c.g.s.), with $T_\mathrm{p}=10^5$ -- $10^7$ and
$n_\mathrm{p}=10^7$ -- $10^{10}$cm$^{-3}$;
\\
\item rotation:
if the star does not rotate, the thermal plasma density is
considered constant within the inner magnetosphere; if, on the contrary,
we consider the rotation, the thermal plasma density decreases linearly
outward, while the temperature increases. $T_\mathrm{p}$ and $n_\mathrm{p}$
are considered as the values at $r=R_\ast$.\\
\end{itemize}

The behaviour of $S_\mathrm{max}$ and $\Delta$
versus the magnetic shell thickness $l$ is investigated for models computed 
assuming different values of $N_\mathrm{e}$, $\delta$, $T_\mathrm{p}$, 
$n_\mathrm{p}$, and rotation.
Fig.~\ref{smz_017} shows some examples of the adopted method
of searching for good combinations of the free parameters.
The shaded areas represent the observed values of $S_\mathrm{max}$
and $\Delta$ derived from the observed 5 GHz lightcurves.
For each value of $N_\mathrm{e}$, the good solutions must show the
observed values of $S_\mathrm{max}$ and $\Delta$ for the same value of
$l/r_\mathrm{r}$. They are reported in Table~\ref{solutions} for the two stars.
%%============================================fig 6
\begin{figure*}
\centering
  \includegraphics[width=17cm]{tr.f6a.ps2}
  \includegraphics[width=17cm]{tr.f6b.ps2}
  \caption[ ]{Simulated brightness distribution of HD\,37479
at $\nu=5$~GHz and $\nu=15$~GHz for different values of
rotational phase. Stellar orientations corresponding to the extremes
and the null effective magnetic field are indicated.}
\label{film1}
\end{figure*}
%%============================================fig 7
\begin{figure*}
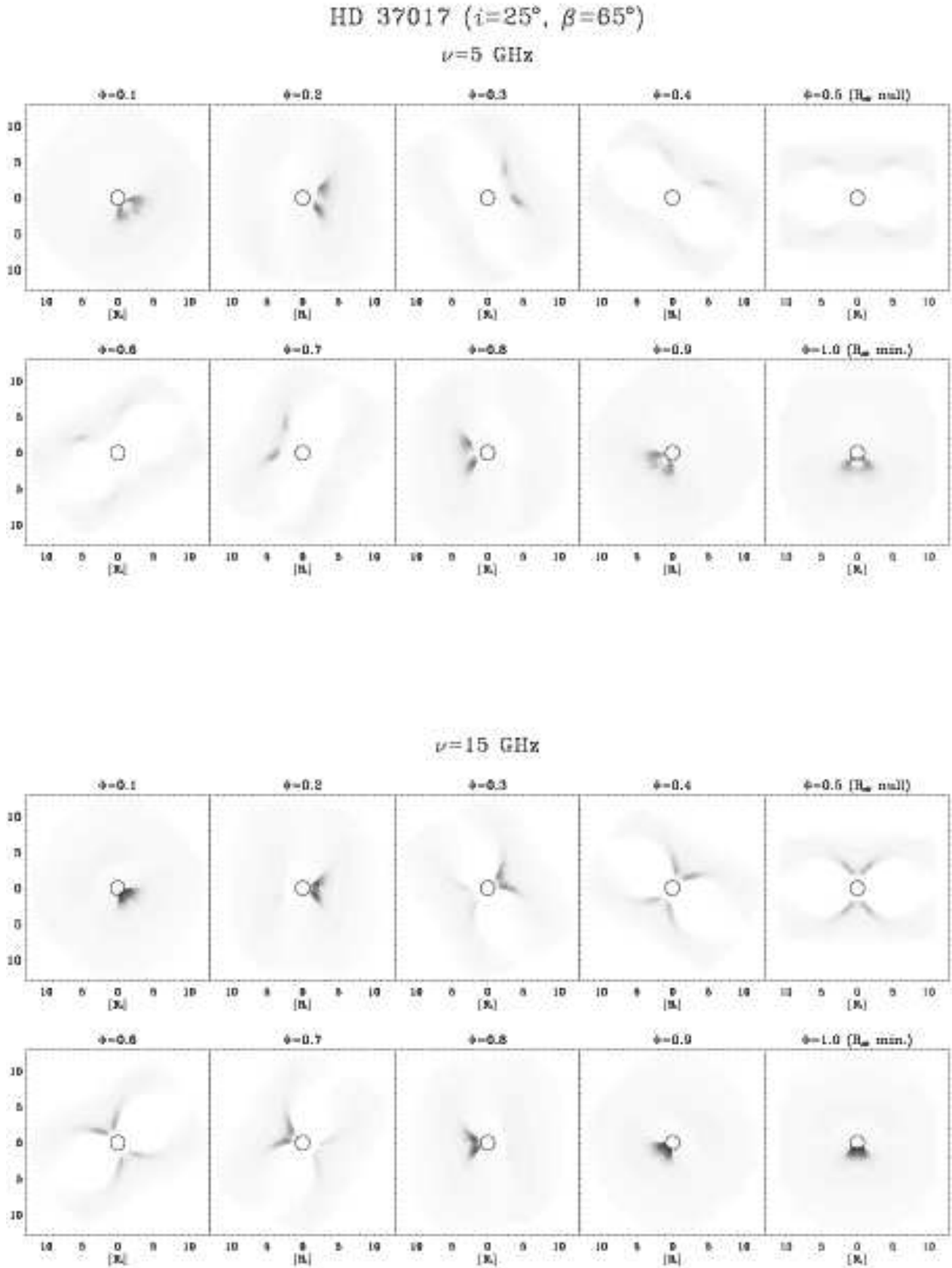

\centering
  \includegraphics[width=17cm]{tr.f7a.ps2}
  \includegraphics[width=17cm]{tr.f7b.ps2}
  \caption[ ]{As Fig.~\ref{film1} for HD\,37071.}
\label{film2}
\end{figure*}
%%======================================================

\section{Discussion}
The model presented in this paper is used to test the scenario proposed
to explain the origin of radio emission from MCP stars
and, if possible, to derive quantitative information about the physical
conditions of their magnetospheres.\\

First, the characteristic quantities $S_\mathrm{max}$ and $\Delta$ of the
lightcurves at 5~GHz of \object{HD\,37479} and \object{HD\,37017}
are computed for different 
combinations of $l$, $N_\mathrm{e}$, $\delta$, $T_\mathrm{p}$, 
$n_\mathrm{p}$, including also the different behaviour of the thermal
plasma in the inner magnetosphere as the star rotates or not.
All the possible solutions are shown in Table~\ref{solutions}.\\

Second, for any possible solution, we compute the emerging radio
flux also at 8.4 and 15~GHz; the spectral index of the simulated spectra
must be flat (say in the range --0.1 -- 0.4) as reported by Drake et al.
(\cite{drake_et_al}) and Leone et al. (\cite{leo96}).
The solutions with the required spectral index are indicated by an
asterisk in the last column of Table~\ref{solutions}.\\
The model is not computed at 1.4 and 22~GHz because:\\
1) at low frequency the emitting region extends far from the star and
close to the Alfv\'en surface, where the geometry of the magnetosphere is not yet 
well known, and\\
2) at high frequency the magnetosphere must be so closely sampled
that computational times are prohibitive.\\

Third, we compute the radio lightcurves at 8.4 and 15~GHz with the aim of
predicting the behaviour of our model, which should be tested in the future with
multi-frequency observations.

\subsection{Effect of the rotation}
The radio lightcurves at 5~GHz for the two stars can be reproduced by
several combinations of free parameters. 
Particularly, the effects of the rotation seem to be important.

Neglecting the effects of rotation, i.e. assuming that the density and the 
temperature of the inner magnetosphere are constant, we find that 
$T_\mathrm{p}=10^5$ K, the electron energy spectrum must be hard ($\delta=2$) 
and a small relativistic electron number
$N_\mathrm{e}$ is required to reproduce the observed radio emission. In this case
$n_\mathrm{p}$ must be quite high ($>1\times 10^9\mathrm{\,cm}^{-3}$),
in order to have the observed modulation of the radio lightcurve.
Also if $T_\mathrm{p}=10^6$ K we get solutions for both stars, for $\delta=2,3,4$;
$N_\mathrm{e}$ increases as the electron energy spectrum becomes softer
($\delta=4$), reaching almost the number density of the Alfv\'en point
(i.e. $\approx 10^6\mathrm{\,cm}^{-3}$).
$n_\mathrm{p}$ is lower than in the previous case, being in the range
$10^8$ -- $10^9\mathrm{\,cm}^{-3}$.\\
For $T_\mathrm{p}=10^7$ K, only the behaviour of \object{HD\,37479} can
be reproduced, but only for $\delta=4$ and very high values of $N_\mathrm{e}$.

The situation is different if the rotation has been taken into account,
considering that $T_\mathrm{p}$ increases linearly outward and
$n_\mathrm{p}$ decreases according to the relation
$T_\mathrm{p}n_\mathrm{p}=const$. In Table~\ref{solutions}
their values extrapolated at $R=R_\ast$ are reported.\\
For $T_\mathrm{p}=10^5$ and $\delta=2,3,4$, the lightcurves of
both stars are well reproduced. The number density
at the base of the inner magnetosphere is about $10^9\mathrm{\,cm}^{-3}$.\\
For $T_\mathrm{p}=10^6$ K we get solutions only for \object{HD\,37479},
only with $\delta=4$ but with $N_\mathrm{e}\approx n_\mathrm{e,A}$.\\
Simulations at $T_\mathrm{p}=10^7$ K do not show 
any possibility for \object{HD\,37017}, while for \object{HD\,37479}
they are limited to $\delta=4$ and very high $N_\mathrm{e}$.
%%============================================fig 8
\begin{figure}
\resizebox{\hsize}{!}{\includegraphics{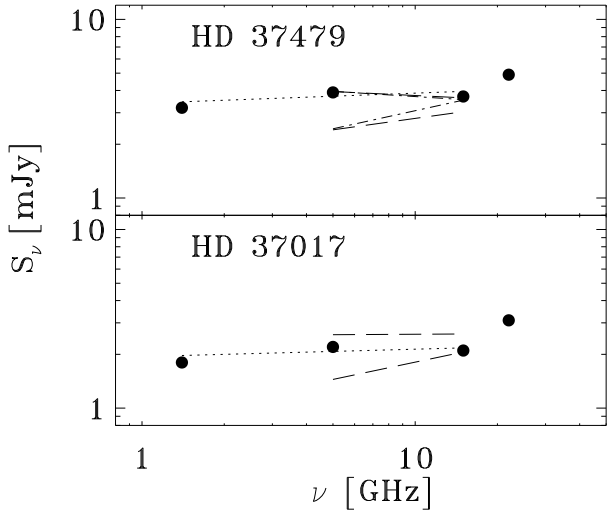}}
 \caption[ ]{
Top panel: Observed spectrum of HD\,37479 by Leone et al. (\cite{leo96})
($\bullet$) and its linear 2 -- 15 GHz fit (dotted lines) are compared with
the computed spectra at minimum and maximum radio emission
for the two solutions marked with asterisks in Table~\ref{solutions}. 
In the order they are listed, 
dashed and dashed-dotted lines represent these solutions.\\
Bottom panel: with the previous symbols, the observed spectrum of  HD\,37017 
is compared with the only solution found.
}
\label{spettri}
\end{figure}

%%============================================fig 9
\begin{figure*}
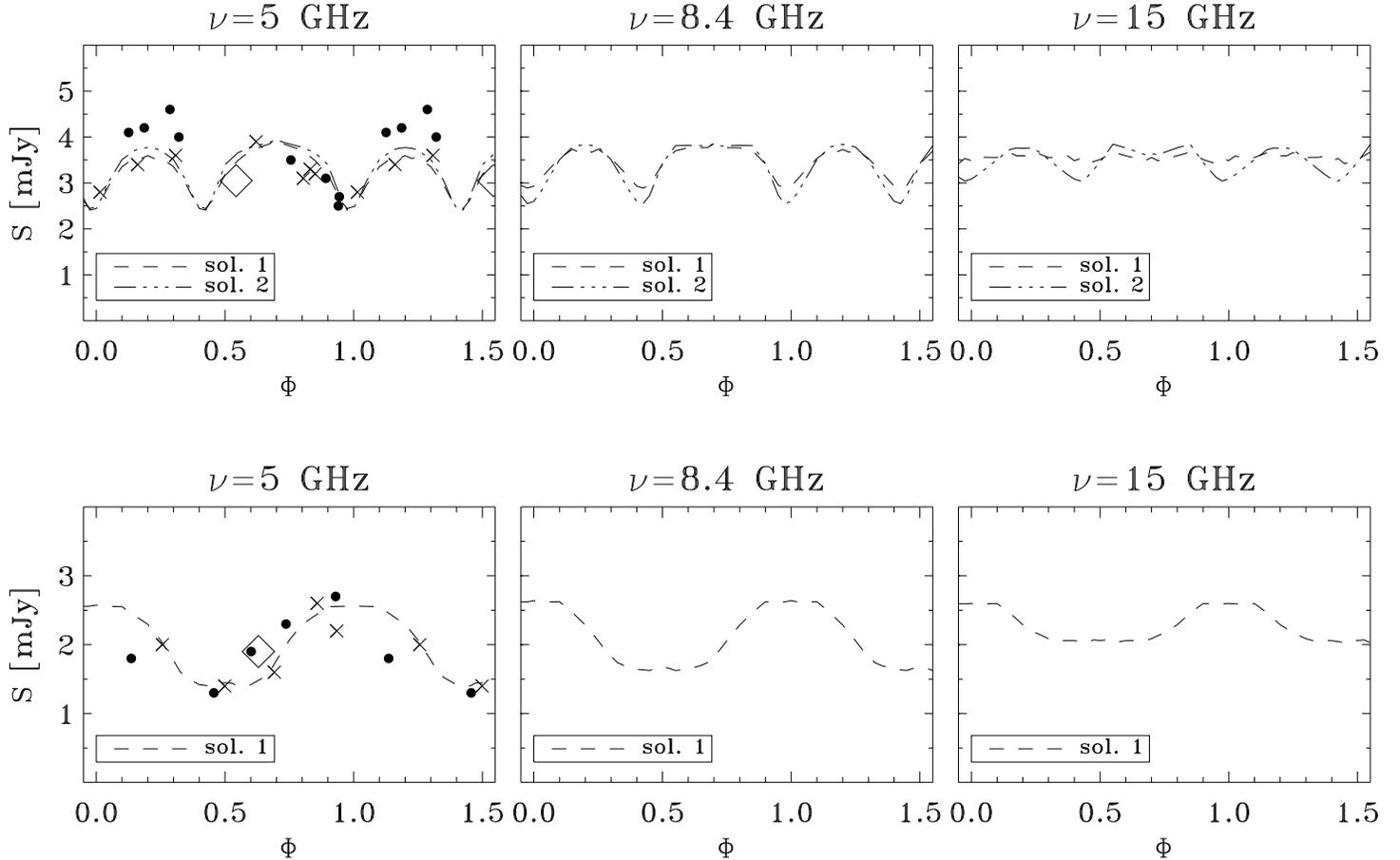

\resizebox{\hsize}{!}{\includegraphics{tr.f9a}}
\resizebox{\hsize}{!}{\includegraphics{tr.f9b}}
 \caption[ ]{
Left panels: 
comparison of the observed 5~GHz data of Fig.~\ref{2cp} and the computed
lightcurves corresponding with the accepted solutions of Table~\ref{solutions}.
Central and right panels: lightcurves at 8.4 and 15\,GHz as predicted by the model.
See Fig.~\ref{2cp} for the meaning of the symbols.
}
\label{fit}
\end{figure*}

\subsection{The effect of the electron energy spectrum}

As already noted, a hard non-thermal electron energy population, such as
$\delta=2$, is able to reproduce the observed radio emission with small
$N_\mathrm{e}$ ($\approx 10^2$ -- $10^3\mathrm{\,cm}^{-3}$). A softer population,
say $\delta=4$, even if able to reproduce the emission, requires a very
efficient acceleration process, as almost all the electrons flowing
with the stellar wind should be accelerated up to relativistic energies.\\
Our results indicate that a hard electron energy population is more
probable.

\subsection{Simulation of radio lightcurves and spectra}

In order to get the spectral index, radio fluxes at 8.4 and 15~GHz are
%have been
computed for any solution at maximum and minimum lightcurve.
The range of variation of the spectral index is also reported in
Table~\ref{solutions}. Only two solutions for \object{HD\,37479} and one
for \object{HD\,37017} give spectral indices similar to the observed ones.
They have in all cases a quite hard electron energy distribution ($\delta=2$)
and are all consistent with a rotating magnetosphere, where the density
of the absorbing matter in the inner magnetosphere decreases outward quite
linearly, while the temperature increases.
Radio spectra at maximum and minimum flux are shown in Fig.~\ref{spettri}
for the accepted solutions, together with the spectral data from
Leone et al. (\cite{leone_et_al}). There is a good agreement, and this result
encourages us to extend the model at lower and higher frequency.

The simulated radio lightcurves for \object{HD\,37479} and \object{HD\,37017},
corresponding to the three accepted solutions, are shown in Fig.~\ref{fit}.
In the left panels, the 5~GHz simulations are superimposed on the measurements
from the literature.
For the first star, there is no significant difference between the two
simulated radio lightcurves and it is, therefore, impossible to discriminate
between them by using only single frequency data.
Even though radio lightcurves are not available for those two stars at other
frequencies, we computed the synthetic curves at 8.4 and 15~GHz in order to give
a prediction of the behaviour in a wide range of frequency.
The comparison between the results of our model computed
with multifrequency observations is therefore crucial.

Both stars are unfortunately too far away (Table~\ref{tab_cp}) to give us the
opportunity to resolve the radio sources by using Very Long Baseline
Interferometry (VLBI). In fact, at 5~GHz, the extent of the radio emitting 
region (Fig.~\ref{film1}) is about 10 stellar radii, corresponding to an
angular size of about 1.5 milliarcseconds (mas).
The angular resolution of the VLBI arrays (with a maximum baseline of about
8000~km) is 1.5~mas, and the source cannot resolved. This is in agreement
with the results of Phillips \& Lestrade et al. (\cite{phill}) who did not
resolve \object{HD\,37479} with high-sensitivity intercontinental VLBI.
At 15~GHz the resolution of the VLBI is 0.5~mas and the
sources can be only barely resolved if observed at a rotational phase
corresponding to the zero of longitudinal magnetic field $B_\mathrm{e}$,
when the radio source has its maximum angular extent (a little less than 10 
stellar radii). A closer MCP star should be observed in VLBI in order to test
the morphology predicted by our model.

\subsection{X-ray emission}
The physical conditions of the inner magnetosphere play an important role
in the emerging radio flux because of the free-free absorption by plasma.
The inner magnetosphere is modelled following
the magnetically wind-shock model by BM97. 
By performing more than 56\,000 simulations per star, 
changing all the free parameters of our model,
we find only 3 solutions satisfying all the observational 
constraints.  
All 3 solutions indicate that the effects
of rotation are not negligible. For the two stars, we find that the 
temperature increases outward from about $10^5$ to $1.4\times 10^6$ K, while the 
density decreases from $2-3\times 10^9$ to $2\times 10^8\mathrm{\,cm^{-3}}$.
As a by-product of our model, we can estimate the X-ray emission of the inner 
magnetosphere once its configuration has been found according to the
radio emission model. Following a procedure similar to that used for the
computation of the radio flux, we compute for each element of the 
magnetosphere the temperature and the number density in order to get the 
Bremsstrahlung emission coefficients for a thermal plasma. 
The emitted power is then 
integrated in the range of energy $0.1-2.0$~KeV, and the resulting X-ray 
luminosities are evaluated as 
$\lg L_\mathrm{X}=30.06$ for \object{HD\,37479} and 
$\lg L_\mathrm{X}=30.42$ for \object{HD\,37017}.
These values are consistent with the limits of detection of the {\em ROSAT} 
all-sky survey, as in Sect.~\ref{appli}.
However, with the new X-ray telescopes now available 
({\em Chandra} and {\em XMM}), both
stars should be easily detectable. In fact, the corresponding fluxes at the
Earth would be respectively $1.2\times 10^{-13}$ and 
$2.4\times 10^{-13}\mathrm{\,erg\,s^{-1}\,cm^{-2}}$, and the detection limits
for a point source 
for the two instruments are $10^{-15}\mathrm{\,erg\,s^{-1}\,cm^{-2}}$
in a few 10\,Ksec.

Recently, Schulz et al. (\cite{schulz}) found that the X-ray emission from the 
MCP-like O7V star \object{$\theta^1$~Ori~C} (Donati et al. \cite{donati2002})
revealed by {\em Chandra}  
is consistent with a magnetically confined wind. Unfortunately, 
radio observations of this object cannot be used to test our model as
this star is embedded in the Orion nebula, one of the brightest radio sources.

The capability to reproduce also the X-ray emission represents an independent and 
further confirmation of the validity of our model.
Multi-wavelength radio observations in conjunction with X-ray
observations will be a real test for our model.

\section{Conclusion}
The study of radio emission from the MCP stars and, in particular, the analysis
of the observed modulations offer a unique opportunity to determine  
the physical properties of the stellar magnetosphere.

The numerical model developed in this paper
is used to reproduce the 5 GHz radio lightcurves
of two well known MCP stars: \object{HD\,37479} and \object{HD\,37017}.

The capability of our numerical model to give us an estimation
of the physical condition of the magnetosphere
is limited by the lack of multifrequency radio lightcurves.
However, the spectral information available up to now allows us to focus
on a small range of possible solutions, all of them indicating a hard energetic
population of non-thermal-emitting electrons, and an inner magnetosphere
filled by a thermal plasma consistent with a wind-shock model that 
provides also X-ray emission.

The possibility to test the radio curves at more than one frequency
would allow us to put more stringent constraints on the model.

Our simulations provide useful information that may be summarized as follows:
\begin{itemize}
\item[ ] we confirm the qualitative model proposed by several authors
to explain the origin of non-thermal electrons as responsible for the observed
radio emission from MCP stars;\\
\item[ ] we point out the importance of the thermal electrons
trapped in the inner magnetosphere for reproducing the rotational modulation of
the measured radio flux density and the radio spectra of the MCP; 
this plasma can prroduce X-ray emission;\\
\item[ ] the length of the current sheets, where the electrons are 
accelerated up to relativistic energies, is about one half of the Alfv\'en
radius; the acceleration process has an efficiency of about $10^{-4}$, as 
only this small fraction of the electrons in the current sheets is accelerated ;
those non thermal electrons have an hard energetic spectrum 
($N(\gamma)\propto (\gamma-1)^{-\delta}$), with $\delta\approx 2$;
the inner magnetosphere is filled by thermal plasma, whose temperature
increases outward from $10^5$ up to $10^6$ K, consistent with a rotating
magnetosphere.
\end{itemize}

\noindent We emphasize the importance of multifrequency radio lightcurves
and contemporaneous X-ray observations
for a further test our model. This will be a powerful
investigation tool for studying the physics of the stellar magnetosphere.

\begin{acknowledgement}
We thank Carlo Nocita for his help in making some figures. We particularly
thank the referee Dr. T. Montmerle for his constructive criticism which
enabled us to strongly improve this paper.
\end{acknowledgement}

%____________
\appendix
\section{}

\subsection{Procedures}
\label{procedure}
We numerically developed the physical
scenario proposed to explain the origin of radio emission arising
from the magnetosphere of MCP stars. %(Sect.~\ref{origin}).
The aim is to reproduce the centimetric radio emission
as a function of the rotational phase.

For any rotational phase and a given wavelength, the procedure is:
\begin{itemize}
\item 3D sampling of the magnetosphere and calculation of the magnetic field 
    vector $\vec{B}$;
\item definition of the Alfv\'en radius and of inner, middle and outer
     magnetosphere;
\item calculation of the number density $n_\mathrm{e}$ of the non-thermal
    electrons in each point of the grid;
\item calculation of emission and absorption coefficients;
\item integration of the transfer equation along paths parallel to the line
    of sight;
\item brightness distribution in the plane of the sky, total flux emitted
    toward the Earth.
\end{itemize}

\subsection{The reference frame}
\label{frame}
In the oblique rotator model the magnetic axis 
has an obliquity $\beta$ with respect to the rotational axis $\vec{\Omega}$,
which in turn forms an angle $i$ with respect to the line of sight.
Thus the orientation of the magnetosphere changes with respect to the line 
of sight as a function of the rotational phase $\Phi$. 
For the purpose of the model,
the most convenient reference frame where to  sample the magnetosphere
is the fixed frame $Ox^{\prime}y^{\prime}z^{\prime}$, with origin $O$
at the center of the star, axis $x^{\prime}$ direct toward the Earth, 
axes $y^{\prime}$ and $z^{\prime}$ in the plane 
of the sky with $z^{\prime}$ coinciding with the projection of $\vec{\Omega}$
on the plane of the sky. 
However, since a dipole has an axial symmetry, the most convenient reference 
frame to compute the vector $\vec{B}$ and all the other physical parameters
is the frame $Oxyz$, anchored to the star, having axis $z$ coinciding with the 
axis of the dipole. Here the axis $x$ lies in the plane defined by the 
rotational and dipole axes. In this frame, the components of vector
$\vec{B}$ are:
\begin{displaymath}
B_{x}= 3m \frac{xz}{r^{5}}
\end{displaymath}
\begin{equation}
B_{y}= 3m \frac{yz}{r^{5}}
\label{campo}
\end{equation}
\begin{displaymath}
B_{z}= m\left( 3 \frac{{z}^2}{r^5}-\frac{1}{r^3}\right)
\end{displaymath}
\noindent
with $m$ the magnetic momentum ($m=\frac{1}{2} B_\mathrm{p} R_{\ast}$).
If we indicate by ${\bf \Re} (\Phi)$ the transformation matrix 
to pass from the frame $Oxyz$ to $Ox^{\prime}y^{\prime}z^{\prime}$,
the magnetic field vector in 
$Ox^{\prime}y^{\prime}z^{\prime}$ is given by:
\begin{equation}
{\vec{B}}^{\prime}({\vec{r}}^{\prime})=\Re \vec{B} (\Re ^{-1} {\vec{r}}^{\prime}).
\label{transform}
\end{equation}

\subsection{Rotation of the magnetosphere}
\label{rotation}
All the physical quantities of the magnetosphere, either vectorial or
scalar, are easily computable in the stellar reference frame $Oxyz$.
However, we need to know them in the reference frame 
$Ox^{\prime}y^{\prime}z^{\prime}$ of the observer. For this purpose, we
need to know the transformation matrix $\Re$ of Eq.(\ref{transform}).

\begin{figure}
\resizebox{\hsize}{!}{\includegraphics{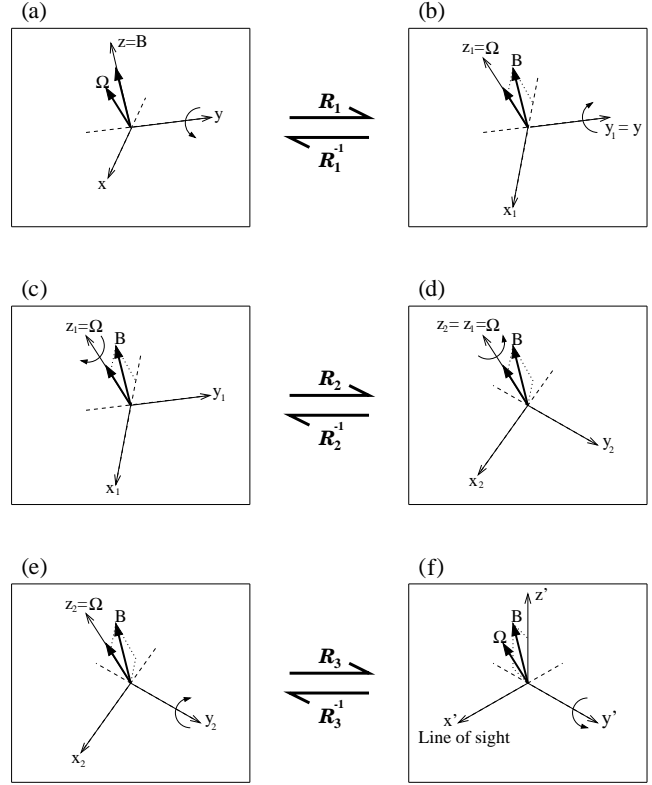}}
 \caption[ ]{Three rotations permit to pass from the stellar reference frame
$Oxyz$ (panel a) to that of the observer $Ox^{\prime}y^{\prime}z^{\prime}$
(panel f).
}
\label{geocp}
\end{figure}

The stellar observing system $Oxyz$ is defined by the 
direction of the magnetic axis ($z$ axis),
and the intersection between the plane of the magnetic equator and the
plane where the $z$ axis and the rotational axis $\Omega$ lie 
(Fig.~\ref{geocp}a).

In the observing reference frame $Ox^{\prime}y^{\prime}z^{\prime}$ the
stellar orientation is univocally defined by the inclination $i$
of the rotation axis with respect to the line of sight, 
the misalignment $\beta$ of the magnetic axis with respect to the rotation 
axis $\Omega$ and the angle $\Phi$ corresponding to the rotational phase.
The three angles $i$, $\beta$ and $\Phi$
define the transformation matrix $\Re(i,\beta,\Phi)$ between 
$Oxyz$ and $Ox^{\prime}y^{\prime}z^{\prime}$.
 
The matrix $\Re$ is the result of three space rotations:
\begin{itemize}
\item[ ]
\begin{eqnarray*}  
\Re_1 =  \left(	 	\begin{array}{ccc}
  \cos \beta	&0	&- \sin \beta	\\
0		&1	&0		\\
  \sin \beta	&0	&\cos \beta
			\end{array} 
\right) 
\end{eqnarray*}
rotation of an angle $\beta$ around the $y$ axis (Fig.~\ref{geocp}~a-b) so that 
$z\rightarrow z_1$ coincides with the rotation axis $\Omega$;\\

\item[ ] 
\begin{eqnarray*}  
\Re_2 =  \left(	 	\begin{array}{ccc}
  \cos \Phi	&-\sin \Phi	&0	\\
  \sin \Phi	& \cos \Phi	&0	\\
  0		&0		&1
			\end{array} 
\right) 
\end{eqnarray*}
rotation of an angle $-\Phi$ around $\Omega$ (Fig.~\ref{geocp}~c-d)
so that $x_1\rightarrow x_2$ lies in the plane defined by the line of sight 
and $\Omega$;\\

\item[ ]
\begin{eqnarray*}  
\Re_3 =  \left(	 	\begin{array}{ccc}
  \sin i	&0	&\cos i	\\
0		&1	&0		\\
 -\cos i	&0	&\sin i
			\end{array} 
\right) 
\end{eqnarray*}
rotation of $(\pi/2-i)$ around $y_2$ axis (Fig.~\ref{geocp}~e-f)
so that $x_2\rightarrow x_3$ coincides with the line of sight 
(axis $x^{\prime}$).
\end{itemize}
The transformation matrix $\Re$ can so be defined as:
\begin{displaymath}
\Re = \Re_3 \Re_2 \Re_1
\end{displaymath}
\noindent The inverse transformation, from $Ox^{\prime}y^{\prime}z^{\prime}$
to $Oxyz$, is then defined by the three inverse rotations:
\begin{displaymath}
\Re^{-1} = {\Re_1}^{-1} {\Re_2}^{-1} {\Re_3}^{-1}
\end{displaymath}
\noindent Rotation matrices are unitary, so the inverse rotations are
described by the transposed matrices.

%%============================================fig A1
%%=======================================================

\section{}
\subsection{The gyrosynchrotron emission}

\label{emission}
\noindent
To calculate the total flux density radiated by the magnetosphere, 
the equation of radiative transfer was numerically integrated,
independently for the ordinary and extraordinary modes of propagation,
along the axis $x^{\prime}$.

For the gyrosynchrotron emission and absorption coefficients 
$\eta_{\nu}$ and $k_{\nu}$, the approximate expressions given by Klein 
(\cite{klein}) were used if the harmonic of gyrofrequency was bigger then 4; 
otherwise the general expressions given by Ramaty (\cite{ramaty}) have been 
adopted.

The presence of the star was taken into account by putting at the center of the 
grid a sphere of radius equal to the stellar radius ($R_{\ast}$), and opaque to the 
radiation.

Defining $i,j$ and $k$ as the index over $x^{\prime}$, $y^{\prime}$ and 
$z^{\prime}$, the method used to integrate the equation of radiative transfer 
may be summarized as follows:
\begin{itemize}
\item calculation of the specific intensity inside each cube element of 
geometrical depth $\Delta l$:
\begin{displaymath}
\Delta I_{\nu}(i,j,k)= 
\frac{\eta_{\nu}(i,j,k)}
{k_{\nu}(i,j,k)}
\left[1-e^{-k_{\nu}(i,j,k)\Delta l}\right]
\end{displaymath}
\item calculation of the optical depth of the column matter between each
grid element and the Earth:
\begin{displaymath}
\tau_{\nu}(i,j,k)= 
\sum_{i^{\prime}=i+1}^{N}
k_{\nu}(i^{\prime},j,k)\Delta l
\end{displaymath}
\item calculation of the specific intensity $I_{\nu}$ arising from each 
direction parallel to line of sight:
\begin{displaymath}
I_{\nu}(j,k)= \sum_{i} \Delta I_{\nu}(i,j,k)e^{-\tau_{\nu}(i,j,k)}.
\end{displaymath}
\end{itemize}
Each element of matrix $I_{\nu}(j,k)$ 
represents the specific intensity in the plane of the sky.

The total flux density can be derived using the relation:
\begin{displaymath}
S_{\nu}=\frac{1}{D^2} \sum_{j} \sum_{k} I_{\nu}(j,k)\Delta l^2
\end{displaymath}
\noindent
where $D$ is the distance of the source.

\end{document}